\documentclass{emulateapj}

\def\saxj{SAX~J1808.4$-$3658}

\def\T0{T^*_0}

\begin{document}

\title{Order in the Chaos: Spin-up and Spin-down during the 2002 Outburst 
of SAX~J1808.4--3658}

\author{L. Burderi\altaffilmark{1}, T. Di Salvo\altaffilmark{2}, M.T. Menna\altaffilmark{3},
A. Riggio\altaffilmark{1,2}, A. Papitto\altaffilmark{3,4}}

\altaffiltext{1}{Universit\`a degli Studi di Cagliari, Dipartimento
di Fisica, SP Monserrato-Sestu, KM 0.7, 09042 Monserrato, Italy;
email: burderi@mporzio.astro.it}
\altaffiltext{2}{Dipartimento di Scienze Fisiche ed Astronomiche,
Universit\`a di Palermo, via Archirafi 36 - 90123 Palermo, Italy;
email: disalvo@fisica.unipa.it}
\altaffiltext{3}{I.N.A.F. - Osservatorio Astronomico di Roma, via Frascati 33,
00040 Monteporzio Catone (Roma), Italy}
\altaffiltext{4}{Dipartimento di Fisica, Universit\'a degli Studi di Roma
''Tor Vergata'', via della Ricerca Scientifica 1, 00133 Roma, Italy}

\begin{abstract}
We present a timing analysis of the 2002 outburst of the accreting millisecond 
pulsar \saxj. A study of the phase delays of the entire pulse profile
shows a behavior that is surprising and difficult to interpret: superposed to a 
general trend, a big jump by about 0.2 in phase is visible, starting at day 14 after
the beginning of the outburst. An analysis of the pulse profile indicates the 
presence of a significant first harmonic. Studying the fundamental and the 
first harmonic separately, we find that the phase delays of the first harmonic
are more regular, with no sign of the jump observed in the fundamental. 
The fitting of the phase delays of the first harmonic with a model which takes
into account the observed exponential decay of the X-ray flux (and therefore of 
the mass accretion rate onto the neutron star) gives important information on the
torque acting on the neutron star during the outburst. We find that the source 
shows spin-up in the first part of the outburst, while a spin-down dominates at
the end. From these results we derive an estimate of the neutron star magnetic 
field strength.
\end{abstract}


\maketitle

\section{Introduction}

In 1998, the idea that neutron stars in Low Mass X-ray Binaries (hereafter LMXBs) 
are spinning at millisecond periods was spectacularly demonstrated by the discovery 
of coherent X-ray pulsations at $\sim 2.5$ ms in \saxj, a transient X-ray source
with an orbital period of 2~hr (Wijnands \& van der Klis 1998).
For almost four years \saxj\ has been considered as a unique object in which some 
peculiarity of the system allowed the detection of the neutron star
spin.  However, in the last few years six other accreting millisecond pulsars have 
been discovered (see Wijnands 2005 for a review): all of them are transient, with
spin periods in the range between 1.7 and 5.4 msec.

\saxj\  
was soon recognized as a weakly magnetized ($<10^{10}$~G), rapidly
rotating (401~Hz) accreting neutron star (Wijnands \& van der Klis 1998; Chakrabarty
\& Morgan 1998). It is usually found in quiescence, where it shows a 
luminosity below $10^{32}$ ergs/s, and exhibits quasi-periodic outbursts, roughly every
two years, with peak luminosity around $10^{36}$ ergs/s. It shows all the phenomenology 
of a typical LMXB (i.e. type-I X-ray bursts and burst oscillations, kHz QPOs, etc.), 
together with coherent pulsations, which makes this system unique.
Although other X-ray millisecond pulsars have been discovered recently, \saxj, the
only accreting millisecond pulsar for which more than one X-ray outburst has been 
observed, still remains the richest laboratory for the study of old accreting 
neutron stars in binary systems.

Still, no detailed timing studies have been published for this source so far.
During the 2002 outburst of \saxj, the persistent accretion-powered X-ray 
pulsations were detected with a fractional r.m.s. amplitude of 3--10\%. The
pulsar spin frequency derived from these data was approximately
400.97521~Hz at the start of the outburst, and a mean spin-down rate of about 
$2\times 10^{-13}$ Hz~s$^{-1}$ has been claimed (Chakrabarty et al.\ 2003, see
also Morgan et al.\ 2003).

In this paper we report the results of the detailed timing analysis of the
2002 outburst of \saxj; we find that the source shows a puzzling behavior 
of the phase delays, which can be interpreted analysing separately 
the fundamental and first harmonic of the pulse profile.

\section{Timing Analysis and Results}

\saxj\ was in outburst in 2002 between October 15 and November 26, and
was extensively observed by Rossi X-ray Timing Explorer (RXTE). We analyse 
here all the available observations of the RXTE data archive taken during 
this outburst.
We mainly use data from the RXTE Proportional Counter Array (PCA, Jahoda et al.\
1996), which consists of five identical gas-filled proportional counter
units (PCUs), with a total effective area of $\sim 6000$ cm$^2$, sensitive in
the energy range between 2 and 60 keV. We used data collected in generic
Events mode, with a time resolution of $125~\mu$s and 64 energy channels.
These files were processed and analyzed using the
FTOOLS v.5.3.1. In order to eliminate the Doppler effects caused by the Earth
and satellite motion, the arrival times of all the events were converted to
barycentric dynamical times at the Solar system barycenter. The position
adopted for the source was that of the proposed radio counterpart ($0''.4$ 
uncertainty, which is compatible with that of the optical counterpart,
Rupen et al.\ 2002; Giles, Hill \& Greenhill 1999). 
The PCA light curve of the 2002 outburst of \saxj\ is shown in Figure~1 (upper 
panel).
For the spectral analysis we also used data from the High-Energy X-ray Timing 
Experiment (HEXTE, $\sim 20-200$ keV energy range, Rothschild et al.\ 1998).

To analyse these data we used the procedure that is extensively described in 
Burderi et al.\ (2006). We corrected the arrival times of all the events for 
the delays caused by the motion of the neutron star in the binary system, 
using the most updated orbital parameters (Papitto et al.\ 2005).
To compute phases of good statistical significance we divided the whole 
observation in intervals of 2 h, approximately the orbital period of the system:
in this way we average over periodic phase residuals, if any, caused by errors
in our assumed orbital parameters (see e.g.\ Galloway et al.\ 2002; Burderi 
et al.\ 2006). 
We epoch folded each interval of data in which the pulsation was significantly 
detected at the spin period of 2.493919760956 ms with respect to the same 
reference epoch, $T_0 = 52562$~MJD, corresponding to the beginning of the 
outburst. 
The fractional part of the phase was obtained by fitting each pulse profile 
with two sinusoids of fixed periods (corresponding to the fundamental, 
with period equal to the spin period, and the first harmonic, with period 
equal to half the spin period, respectively). In this way we obtain the 
phase delays shown in Figure~2 for the fundamental (right panel) and first 
harmonic (left panel), respectively. 

\begin{figure}
\plotone{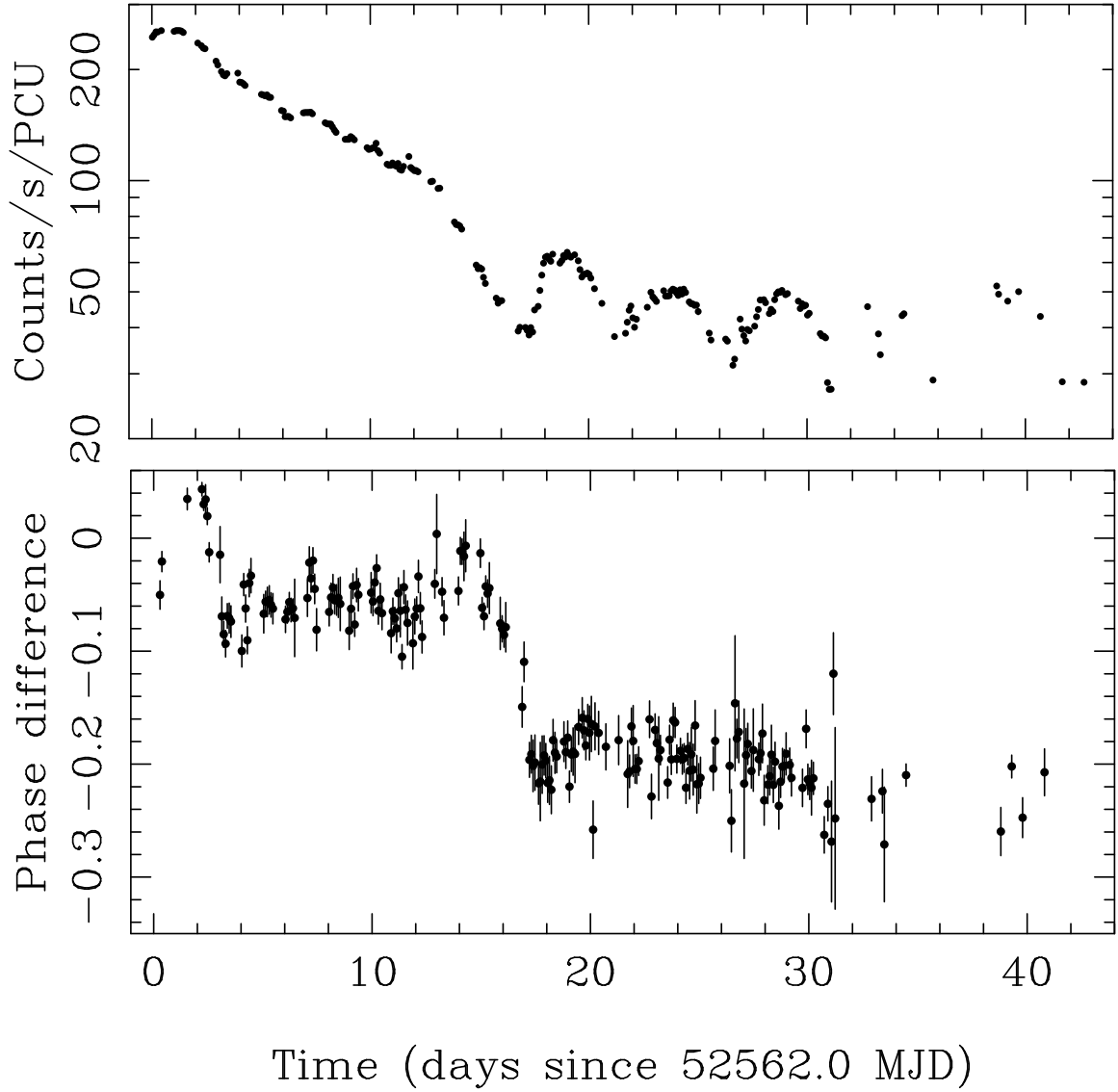}
    \caption{{\bf Upper:} PCA count rate of \saxj\ during the 2002 outburst. The start
    time, $T_0$, is 52562 MJD, corresponding to October 15 2002.
    {\bf Lower:} Phase differences, fundamental minus harmonic, vs.\ time. The jump
    of the phase of the fundamental at day 14 is clearly visible in the differences. 
    \label{fig1} }
\end{figure}

The phase delays of the fundamental show a puzzling behavior, with a 
clear jump by about 0.2 in phase which occurs at day 14 after the beginning 
of the outburst. Interestingly, in the outburst light curve, the 14th day 
corresponds to a change of the steepness of the exponential flux decay 
(see Fig.~1, upper): while before this day the X-ray flux decreases exponentially 
with a characteristic time of about 10 days, after this day the characteristic 
time of the exponential decay becomes about 3 days (a similar behavior of the
light curve was observed for the 1998 outburst of \saxj, Gilfanov et al.\ 1998). 
After day 17 from the beginning of the outburst the X-ray flux shows rapid
oscillations of the count rate on timescales of hours to days.
On the other hand, the phase delays of the first harmonic do not show any
evidence of such a jump. Note that this is not an effect of the larger 
error bars of the phases of the harmonic, as it can be seen from Figure~\ref{fig1}
(lower panel), where we plot the differences between the phases of the 
fundamental and the phases of the harmonic as a function of time. 
For a global shift of the pulse profile the phases of the fundamental and 
harmonic should shift in the same way and the phase differences should remain 
constant. On the other hand, a variation of the differences indicates a shift 
of the phase of the fundamental with respect to the harmonic caused by a change 
of the shape of the pulse profile. 
In the plot of the phase differences it is possible to see that 
the big jump at day 14 is still visible in the differences, and is much 
larger than the associated error bars, meaning that it is present in the 
fundamental and not in the harmonic. We conclude that in this case the 
behavior of the harmonic is genuinely simpler than the behavior of the 
fundamental, and therefore we tried to fit it to an appropriate model. 

In order to derive the differential correction to the spin frequency,
$\Delta \nu_0$, and the spin frequency derivative, $\dot \nu$, we have to
derive a functional form for the time dependence of the phase delays. For a
constant variation of the spin frequency ($\dot \nu = \rm const$), the phase 
delays (which are defined as the double integration over time of $\dot \nu$)
will be a parabolic function of time. However, since for \saxj\ 
the X-ray flux is observed to vary with time, we expect a more complex 
dependence of time. To derive this expression, we use the following simple 
assumptions:
i) The bolometric X-ray luminosity $L$ is a good tracer of the mass
accretion rate, $\dot M$, {\it via} the relation $L = (GM/R) \dot M$, where 
$G$, $M$, and $R$ are the Gravitational constant and the neutron star mass 
and radius, respectively. Therefore, we assume for the mass accretion rate
the following dependence on time: $\dot M (t) = \dot M_0 \exp[(t-T_0)/\tau]$,
where $\tau$, the characteristic time of the flux exponential decay, that is
9.27 days from a fit of the X-ray light curve. 
ii) The accreted matter transfers to the neutron star its specific angular
momentum at the corotation radius (see e.g.\ Rappaport et al.\ 2004). 
In general, the matter should transfer to the neutron star its specific
angular momentum at the accretion radius (that is the magnetospheric radius,
the radius at which the Keplerian accretion disk is truncated by the neutron 
star magnetic field).
Since for accreting millisecond pulsars, the magnetospheric radius must be 
very close to the corotation radius (about 3 neutron star radii in the case of
\saxj), our assumption can be considered a good approximation, and, in any case, 
gives a lower limit to the mass accretion rate needed to obtain the observed 
torque. 
Therefore, the rate of angular momentum transferred to the neutron star is 
$\dot L = 2 \pi I \dot \nu = \dot M (G M R_{\rm CO})^{1/2}$,
where $I$ is the neutron star moment of inertia, $\nu$ its spin frequency,
and $ R_{\rm CO}$ the corotation radius.
We do not consider here any form of threading of the accretion disk by
the magnetic field of the neutron star (see e.g.\ 
Rappaport et al.\ 2004).

From these assumptions, with some algebra, we derive the following expression
for the phase delays caused by accretion torque due to an $\dot M$ decreasing
exponentially with time:
\begin{equation}
\label{phaset}
\phi (t) = \phi_0 - B (t-T_0) - C \exp[-(t-T_0)/\tau]
\end{equation}
where $\phi_0$ is a constant, all the times are expressed in days, 
$C = 1.067 \times 10^{-4} I_{45}^{-1} P_{-3}^{1/3} m^{2/3} \tau^2 \dot m_0$, 
where $I_{45}$ is the moment of inertia in units of $10^{45}$ g cm$^2$, $P_{-3}$
is the spin period in msec, $m$ is the neutron star mass in Solar masses, and
$\dot m_0$ is the accretion rate at $t=T_0$ in units of $10^{-10}$ M$_\odot$ 
yr$^{-1}$, and, finally, $B = \Delta \nu_0 + C/\tau$, where $\Delta \nu_0$ 
is the differential correction to the spin frequency adopted for the folding
of the light curves. Note that we are neglecting here the change to a 
steeper fall off of the light curve which occurs at day 14.
However, as it will become clear in the following, after day 14 the mass 
accretion rate is already so low that any spin-up torque becomes negligible 
(with respect to the behavior of the phase delays which instead tend to flattens,
see below). Therefore, including a reduction of $\tau$ in the model after day 14 
(which further reduces the already negligible spin-up), does not change the results.

We tried to fit the phase delays of the first harmonic with eq.~(\ref{phaset}),
but we obtained a poor description of the data, corresponding to a 
$\chi^2 / dof = 613.1/199$.\footnote{Note that fitting these data to simple linear
or parabolic functions gives similar (slightly worse) results, corresponding to
$\chi^2/dof = 623.4/200$ and $\chi^2/dof = 622.4/199$, respectively.}
Indeed, using this model, we can obtain a good fit of the first 14 days of the 
outburst, but, with respect to this fit, we observe a flattening of the phase 
delays after day 14. To describe this flattening we added to the model described 
by eq.~(\ref{phaset})
a quadratic term corresponding to a constant spin down. In this way we obtained a
significant improvement of the fit ($\chi^2/dof = 485.0/198$; this $\chi^2$ is still 
large, but now we do not see any systematic trend in the residuals with respect to 
the best fit model). Although in this way we have obtained a great improvement of
the quality of the fit, the obtained $\chi^2$ is still unacceptable, because of
localised residuals around day 4 and day $30-31$ from the beginning of the outburst.
These residuals cannot be taken into account modifying the fitting function, but
the uncertainties derived from a fit that is not acceptable in the statistical sense
may be underestimated. To obtain more conservative uncertainties for our best fit
parameters, we therefore increased by a factor 1.5 the errors of all our phase points,
obtaining in this way a $\chi^2 / dof = 215 / 198$ (that is now very close to $1$),
and we have re-calculated all the uncertainties on our best fit parameters.
The best fit parameters and the corresponding uncertainties (at $90 \%$ c.l.) are 
shown in Table~1; the best fit function is plotted on top of the data in Figure~2. 
For a comparison, we have  plotted the best fit function of the first harmonic on 
top of the fundamental.

The last issue we have to discuss regards the systematic uncertainties induced
by the uncertainty on the source position. As discussed in Burderi et al.\ (2006),
these systematics give both a systematic uncertainty on the linear term (that is 
the spin frequency correction) and on the quadratic term (that is the spin 
frequency derivative). For an error circle of $0''.4$ around the position of \saxj,
we find that the systematic uncertainty on this two terms are, respectively,
$\Delta \nu_{\rm sys} = 7.9 \times 10^{-8}$ Hz and $\Delta \dot \nu_{\rm sys} = 
1.8 \times 10^{-14}$ Hz s$^{-1}$. This increases the total uncertainty on the spin 
frequency to $1.1 \times 10^{-7}$ Hz, and the total uncertainty on the spin-up
and spin-down terms, which become: $\dot \nu_0 = (4.40 \pm 0.83) \times 10^{-13}$ 
Hz s$^{-1}$ and $\dot \nu_{\rm sd} = -(7.6 \pm 2.3) \times 10^{-14}$ Hz s$^{-1}$,
respectively. Even considering these systematic uncertainties, both the spin-up and 
spin-down terms are still significant at more than $4 \sigma$ (note that all 
the reported uncertainty are at $90\%$ c.l.).


Finally, in order to compare the derived spin-frequency derivative with the mass
accretion rate $\dot M_0$ as inferred from the bolometric X-ray flux of the source,
we performed a spectral analysis of the PCA and HEXTE spectra at the beginning of
the outburst (i.e.\ at Oct 15 and 16, 2002). Keeping the absorption to the source
fixed at $N_H = 3.0 \times 10^{21}$ cm$^{-2}$ (the Galactic X-ray absorption in the
direction of \saxj\ at a distance of 3.5 kpc, Galloway \& Cumming 2006), a very good 
fit to the $2.5-200$ keV spectrum of \saxj\ is given by a disk blackbody plus a 
cutoff power law, and a Gaussian emission line with centroid fixed at 6.4 keV 
(the fluorescence K-shell iron line). We extrapolated the observed X-ray flux in 
a broad band energy range in order to evaluate the bolometric accretion luminosity, 
which resulted to be $\sim 1 \times 10^{37}$ ergs/s on 2002 Oct 15 assuming a distance 
of 3.5 kpc and correcting for the interstellar absorption.

\section{Discussion and Conclusions}

The results described in the previous section demonstrate that the timing analysis
for an accreting pulsar is more complex than expected. The phase delays of the
fundamental clearly show a jump by 0.2 in phase which may be due to instabilities
induced by the accretion of matter onto a weakly magnetized star. Indeed, if the
magnetic field is not strong enough to completely dominate the motion of matter to
the polar caps, we can expect that variations in the accretion flow may cause small
movements of the footpoint on the neutron star which can give rise to the shift in 
phase that we observe. On the other hand the phase delays of the first harmonic
show a more regular behavior and can be fitted to a model derived from the theory of
the accretion torque, as it will become clear in a while. Therefore, the first
harmonic might have some fundamental physical meaning, probably related to the fact
that this may represent the accretion onto both the polar caps.

As mentioned above, the results of the fitting of the phase delays of the 
harmonic are in agreement with torque onto the neutron star predicted by the
accretion theory. A simple second order polynomial 
or a model taking into account the exponential decrease with time of $\dot M$
do not give a good fit of the phase delays in the whole time range;
these models give a good fit if we consider only the first 15 days of the outburst. 
We obtain a good fit of the phase delays during the whole outburst
using the model described by eq.~\ref{phaset} and adding 
to this model a quadratic term, which describes the flattening of 
the phases at the end of the outburst. This means that the neutron star is 
spinning-up at the beginning of the outburst, as expected when the mass accretion 
rate is relatively high, but spins-down at the end of the outburst. 

This gives very important information on
the torque acting onto the neutron star in this system, implying that, when the
mass accretion rate has significantly decreased, the torque onto the neutron star
changes sign. This can be caused, for instance, by a threading of the accretion 
disk which becomes important at low accretion rate (see eq.~(23) in Rappaport et
al.\ 2004). In this case, we can evaluate the magnetic field of \saxj\ from
our measured value of the spin-down using the relation: $\mu^2 / (9 r_{\rm CO}^3)
= 2 \pi I \dot \nu_{\rm sd}$. The magnetic field found in this way is 
$B \sim (3.5 \pm 0.5) \times 10^8$ Gauss, perfectly in agreement with previous 
constraints (see e.g.\ Psaltis \& Chakarabarty 1999; Di Salvo \& Burderi 2003).

On the other hand, the spin-up observed in the first part of the outburst will
give information on the mass accretion rate at the reference time $t=T_0$, that
is $\dot M_0 \simeq 1.8 \times 10^{-9}$ M$_\odot$ yr$^{-1}$. This gives a 
bolometric X-ray luminosity at the beginning of the outburst of $\sim 2 \times 
10^{37}$ ergs/s. This is about a factor of 2 higher than the bolometric X-ray
luminosity inferred by the \saxj\ RXTE spectra (about $1 \times 10^{37}$ ergs/s
assuming a distance of 3.5~kpc). On the other hand, assuming that the mass accretion 
rate, $\dot M_0$, inferred from our timing analysis is correct, from a comparison with 
the observed X-ray flux, we can conclude that the source should be at a distance 
of $\sim 5$~kpc.

\acknowledgements
This work was partially supported by the Ministero della Istruzione,
della Universit\`a e della Ricerca (MIUR).



\begin{deluxetable}{lclc}
\tablecaption{Best fit parameters of the phase delays of the first harmonic. }
\tablehead{
\colhead{Parameter} &
\colhead{Value} &
\colhead{Parameter} &
\colhead{Value}
}
\startdata
$B$ & $(-0.8 \pm 2.6) \times 10^{-3}$ & $\Delta \nu_0$ & $-(3.61 \pm 0.72) \times 10^{-7}$ Hz \\
$C$ & $0.282 \pm 0.054$ & $\dot \nu_0$ & $(4.40 \pm 0.81) \times 10^{-13}$ Hz s$^{-1}$ \\
$D$ & $(2.85 \pm 0.59) \times 10^{-4}$ & $\dot \nu_{\rm sd}$ & $-(7.6 \pm 1.5) 
\times 10^{-14}$ Hz s$^{-1}$ \\
\hline
$\dot M_0$ & $(1.81 \pm 0.32) \times 10^{-9}$ M$_\odot$ yr $^{-1}$  & $\nu_0$ & $400.975209690 
\pm 7.2 \times 10^{-8}$ Hz  \\
\enddata
\tablecomments{The best fit function is given by eq.~(\ref{phaset}) + $D (t-T_0)^2$,
where $D = (1/2) \dot \nu_{\rm sd}$. The reference time $T_0$ is the beginning 
of the observation, that is 52562 MJD. Quoted errors are at $90\%$ c.l.\ and
do not include the systematic uncertainties discussed in the text. }
\label{table1}
\end{deluxetable}


\begin{figure}
\plottwo{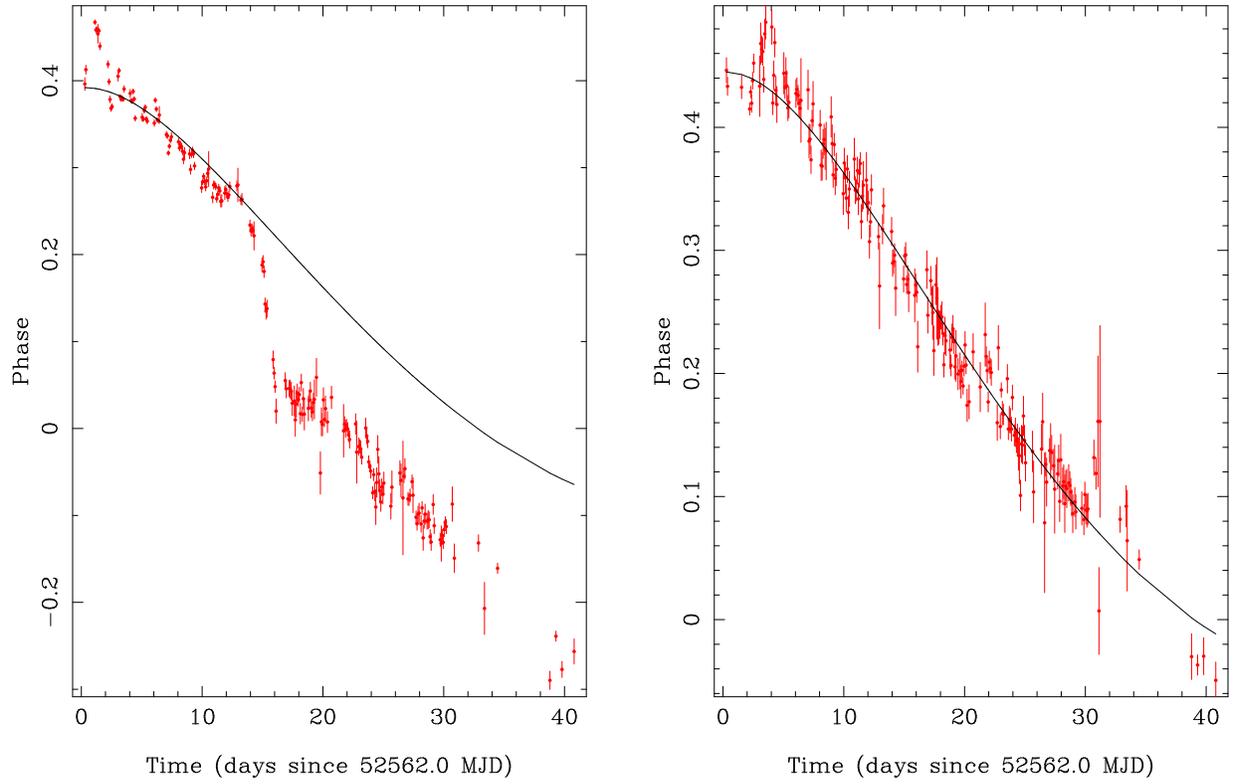}{f2b.eps} 
    \caption{{\bf Left:} Phase vs.\ time for the fundamental of the pulse frequency
    of \saxj. {\bf Right:} Phase vs.\ time for the first harmonic of the pulse 
    frequency of \saxj. On top of the data, the best fit function (the sum of 
    eq.~(\ref{phaset}) and a quadratic term) is plotted as a solid line.
    \label{fig2} } 
\end{figure}

\end{document}